\documentclass[twocolumn,english,aps,superscriptaddress,prl]{revtex4-1}

\usepackage{array}
\usepackage[latin9]{inputenc}
\usepackage{amsmath}
\usepackage{amssymb}
\usepackage{graphicx}
\usepackage{babel}
\usepackage{mathrsfs}
\usepackage{amsfonts}
\usepackage{epstopdf}
\usepackage{multirow}
\usepackage{color}
\usepackage{natbib}

\begin{document}
\title{Three-Dimensional Low-Density Fermions in Planckian Limit:\\
Entropy and Dynamics}

\author{Shao-Jian Jiang}
\affiliation{State Key Laboratory of Magnetic Resonance and Atomic and Molecular Physics,
Wuhan Institute of Physics and Mathematics, APM,
Chinese Academy of Sciences, Wuhan 430071, China}
\author{Fei Zhou}
\affiliation{Department of Physics and Astronomy, University of British Columbia, 6224 Agricultural Road, Vancouver, BC, V6T 1Z1, Canada}

\begin{abstract}
In this Letter, we explore dynamics in a three-dimensional strongly interacting liquid.
In quantum liquids discussed below, thermodynamic properties 
such as pressure and thermal energies are {\em fully} characterized by $s(T)$, the entropy density of the liquid (that is also directly proportional to the hydrodynamic viscosity). 
We obtain a universal fermion spectral function $A(\omega, {\bf k})$ that is distinctly specified by $\hbar / T$, a {\em  Planckian } time scale.
These phenomena can emerge in strongly interacting many-body states with a finite fermion density $\rho$ at temperatures $T^*$ where the chemical potential of fermions $\mu(\rho, T=T^*)$ approaches zero and can be thought as many-body simulations of certain aspects of Planckian dynamics.
\end{abstract}

\date{\today}
\maketitle

Over the last few decades or so, there had been enormous growing interests in dynamics in strongly interacting regimes~\cite{Coleman05,Zhu02,Senthil04,Sachdev10,Liu11,Sachdev11b,Kitaev15,Sachdev93,Maldacena16,Polchinski16,Zaanen04}.
Relatively recent observation of Planckian transport in low dimensional condensed matter systems, where the transport time  was found to be near the value of $\hbar/T$ ($T$ being the temperature),
further indicates universal dynamics in a class of very different strongly interacting systems available in labs~\cite{Legros19,Sarkar18,Moll16,Cao18,Gallagher19,Crossno16}. The interaction time inferred in experiments appears to suggest an intimate and surprising connection between quantum dynamics in a variety of strongly interacting systems~\cite{Damle97,Coleman05,Zaanen04,Sachdev11b} and fastest quantum information spreading near a black hole~\cite{Kitaev15,Maldacena16,Polchinski16}.
These intensive experimental efforts and impressive successes further demand more new theoretical efforts on quantum dynamics in strongly interacting regimes.

Despite all the challenges, there had already been very remarkable theoretical progress made on strongly interacting dynamics. 
For a wide class of {\em low dimensional} quantum critical states, we expect that electron interaction time scales as $\hbar/T$, a hallmark result of quantum criticality~\cite{Sachdev11b}.
This peculiar aspect can also often naturally appear in non-Fermi liquids, either critical or as a stable phase~\cite{Zhu02,Sachdev93,Sachdev10}. It has also been speculated to contribute to the linear-$T$ behavior of resistivity in doped cuprates, a very fascinating phenomena so far only observed in a few very strongly interacting {\em low} dimensional systems~\cite{Zaanen04, Coleman05}.
On strongly interacting systems, general conjectures have also been made on lower bounds of hydrodynamic viscosity in quantum systems based on applications of
the correspondence principle in AdS/CFT~\cite{Kovtun05}.
All these strongly interacting dynamical phenomena had been anticipated to be bounded or limited by $\hbar$, the Planck constant.
Hydrodynamic flows have been recently observed for electrons in ultra clean two-dimensional graphene~\cite{Ku19,Levitov16,Bandurin16,Torre15,Lucas16,Hartnoll07,Andreev11}
and earlier for charge neutral atoms in cold atomic gases~\cite{OHara02,Taylor12,Enss11}.

In this Letter, we explore Planckian dynamics, in {\em three-dimensional low-density attractive fermions instead of low-dimensional electrons that had been previously focused on}, especially dynamic spectral function
$A(\omega, {\bf k})$. Experimentally, the spectral function can be directly studied by the approach of {\em angle resolved photon emission spectroscopy}.
The model we want to focus on is for fermions with $SU(2)$ spin rotation invariant {\em local} interactions,
\begin{widetext}
\begin{eqnarray}
 {\mathcal H} &=& {\mathcal H}_0-\mu N +{\mathcal H}_I, {\mathcal H}_0-\mu N =\sum_{\alpha} \int d{\bf r} \psi^\dagger_\alpha \left(-\frac{\nabla^2}{2} -\mu \right)\psi_\alpha, \nonumber \\
 {\mathcal H}_I&=&\sum_{\alpha} \int d{\bf r} g_{s\alpha}{ \psi^\dagger}_\alpha \sigma_y {\psi^\dagger}^T_\alpha \psi^T_\alpha \sigma_y  \psi_\alpha
+\sum_{\beta \neq\alpha}  \int d{\bf r} ( g_{s3} { \psi^\dagger}_\alpha \sigma_y {\psi^\dagger}^T_\beta \psi^T_\beta \sigma_y  \psi_\alpha
+g_{t}{ \psi^\dagger}_\alpha \sigma_y {\bf \sigma} {\psi^\dagger}^T_\beta  \cdot \psi^T_\beta {\bf \sigma} \sigma_y  \psi_\alpha ),
\label{4fQFT}
\end{eqnarray}
\end{widetext}
where $\psi^T_\alpha=(\psi_{\uparrow,\alpha}, \psi_{\downarrow,\alpha})$ is the field operator for {\em spinful} fermions defined at valley $\alpha=1,2$ and $\mu$ is the chemical potential controlling the density of fermions in a liquid. $g_{s1,2}, g_{s3}, g_{t}$ are four interaction constants for intra-valley spin singlet, inter-valley spin singlet and inter-valley spin triplet channels, respectively.
The four-fermion operators have been ordered in a way most convenient for the rest of discussions. Below we will illustrate that 
Eq.~(\ref{4fQFT}) can lead to {\em three-dimensional} Planckian dynamics that so far had only been directly observed in highly complex strongly interacting 
 {\em two-dimensional} cuprates and a few other two-dimensional electronic systems~\cite{Legros19,Sarkar18,Moll16,Cao18,Gallagher19,Crossno16}.

There can be at least three physical realizations of this simple model.
The most obvious ones, when we set $\alpha=\beta=1$ and take the single valley limit, are cold gases near Feshbach resonances.
Most previous efforts on resonant gases mainly focused on properties of superfluids at low temperatures.
Our proposed Planckian liquids can be related to cold gases near Feshbach resonances but they are not low-temperature superfluids. Instead, they correspond to strongly interacting non-superfluid {\em quantum matter} above critical temperature $T_c$ of superfluids. Their unique equation of states is also qualitatively different from that of high-temperature thermal gases and we will return to this more explicitly later when discussing our results.
Other physical systems where Eq.~(\ref{4fQFT}) can be thought as an effective field theory are optical lattices that can simulate the Hubbard-$U$ model but with 
attractive atomic interactions~\cite{Bloch08}.
Because of the tunability of the bandwidth $W$ or $U$ in optical lattices, the negative-$U$ in principle can be larger or even much larger than $W$, effectively leading to flat-band physics which is typically strongly interacting.  And if the density of fermions is relatively low and fermions mainly concentrate near the bottom of a band, then
the low-energy physics can be effectively characterized by Eq.~(\ref{4fQFT}). To our knowledge, this specific strongly interacting limit had not been fully explored in optical lattices.

Perhaps more interestingly, the model can also describe interacting electrons near a Lifshitz transition where chemical potentials approach bottoms of a conduction band (with either single or multiple valleys) from below and
fermions have an emergent quadratic dispersion in the low-density limit. Previously studied low-density strongly interacting electrons~\cite{Lin13} 
could be in such a class of experimental systems where fascinating dynamics can be potentially studied.
Typically, in low-density electrons, Coulomb interactions are much stronger than phonon
mediated interactions, unlike in conventional metals where, with Fermi energies much higher than the Debye frequencies,
Coulomb interactions are further logarithmically suppressed near Fermi surfaces. 
However, recent experiments on low-density systems seem to suggest otherwise in both three dimensions (3D)~\cite{Lin13} and two dimensions (2D)~\cite{Cao18,Park20,Hao20}. 
Theoretical progress based on the general idea of Kohn-Luttinger mechanism
suggests that repulsive interactions can effectively induce attractive ones and various pairing phenomena in different electronic channels~\cite{Raghu10,Nandkishore14,Crepel21,Slagle20,Kohn65}.
They can even become strongly coupling when on-site repulsion $U$ is of the order of the band width $W$.
For instance, it was proposed quite recently that an effective theory of {\em spinless} repulsive $f$-band electrons in a honeycomb lattice~\cite{Crepel21} can be cast in the form of Eq.~(\ref{4fQFT}) in the spinless limit.

To visualize scale symmetry in this model, we carry out standard scale transformations and implement them using conventional renormalization group equations (RGEs).
In $d$ dimensions near $\mu=0$, the Hamiltonian
${\mathcal H}(\tilde{g}, \tilde{\mu}; \Lambda)$
defined at an ultraviolet (UV) momentum scale $\Lambda$ then transforms into a new one defined at a new scale $\Lambda + d \Lambda$ 
following the equation below;

 \begin{eqnarray}
 && \frac{d\tilde{\mu}}{ds} =\beta_\mu=-2 \tilde{\mu}, \frac{d{Z}}{ds} =\beta_Z= 0,
  \nonumber \\
 && \frac{d\tilde{g_i}}{ds}=\beta_{g_i}=(d-z)\tilde{g_i} +\tilde{g_i}^2, i=s1,s2, s3, t.
 \label{RGE}
 \end{eqnarray}
 where $s= \ln \Lambda$
and $Z$ describes the fermion field renormalization.
We have introduced dimensionless coupling  constants $\tilde{g}_i=C_i g_i\Lambda^{d-z}$ ($C_i$ are constants of order unity) and chemical potential $\tilde{\mu}=\mu \Lambda^{-z}$.
$z=2$ is the temporal scaling exponent.
The RGEs become exact near $\mu=T=0$ or in the low-density limit~\cite{mu};
because of $SU(2)$ spin rotation invariance, we find that $\beta_{g_i}$ for different channels are naturally decoupled.

Eq.~(\ref{RGE}) defines a family of Hamiltonians at different UV scales $\Lambda$ that otherwise are completely equivalent for discussions of infrared physics.
For three dimensions ($d=3$) we are currently interested in,
there are {\em 16 fixed-point solutions} to Eq.~(\ref{RGE}) where the right hand sides of the RGEs vanish.
They correspond to scale invariant Hamiltonians we are interested in and can be classified into four groups:
i) non-interacting one with $g_i=0$, $i=s1,s2,s3,t$ (1); ii) one of the channels is strongly interacting and the other three are non-interacting or vice versa ($4+4$);
iii) two of the channels are strongly interacting and the other two are non-interacting ($6$); iv) all four channels are strongly interacting (1).  In the rest of the Letter, we will discuss
the physics of strongly interacting class ii); classes iii) and iv) are less generic as they require simultaneously fine tuning more than two interaction channels although many discussions below can be easily generalized to those more peculiar situations.

The free-fermion fixed point is specified by $\tilde{\mu}^*=\tilde{g}^*_{i}=0$ and class ii) is characterized by 
\begin{eqnarray}
\tilde{\mu}^*=0, \tilde{g}^*=-(d-2) & & \mbox{  for $d>2$ including $d=3$,}
\label{fp}
\end{eqnarray}
where we only show interactions in the channel that is strongly interacting and have muted the other three channels.
To simplify the presentation, we further take the single valley limit of $\alpha=\beta=1$ without losing generality and set $\tilde{g}=\tilde{g}_{s1}=\tilde{g}^*$.
This strong coupling fixed point exhibits $SO(2,1)$ conformal symmetry~\cite{Hagen72,Niederer72,Nishida07}. 
It can be applied to study strongly interacting electrons in lattices~\cite{Crepel21,Lin13}, or
identified with Feshbach resonances in cold gases~\cite{Werner06,Taylor12,Enss11,OHara02}.

A strong coupling fixed point with  $SO(2,1)$ symmetry~\cite{Hagen72,Niederer72} can be conveniently characterized by correlation functions that exhibit distinct scaling properties because of anomalous scaling dimensions of
various local operators~\cite{Nishida07}. In our case, consider 
simple four-point {\em time-ordered}  correlation functions, 
$G_{4f} ({\bf r}, t)=\langle0|{\mathcal T} \psi^T({\bf r}, t) \sigma_y \psi({\bf r}, t)$ $ \psi^\dagger (0, 0) \sigma_y {\psi^\dagger}^{T}(0,0)|0\rangle$.
It takes a generic form at an $SO(2,1)$ fixed point when $\tilde{g}=\tilde{g}^*$,
\begin{eqnarray} 
G_{4f} ({\bf r}, t > 0) \sim  \frac{1}{t^{\Delta_{4f}/2}}\exp \left(i\frac{{\bf r}^2}{4t} \right),
\label{4fC}
\end{eqnarray}
where $\Delta_{4f}$ features $SO(2,1)$ symmetric fixed point physics. 
$\Delta_{4f}$ in Eq.~(\ref{4fC}) can be directly related to scaling dimensions of {\em four-fermion} operators~\cite{4f}.

At the fixed point of $\tilde{g}^*=2-d$($4>d>2$), following the above approach, one finds that $\Delta_{4f}=4$ while the scaling dimension of two-fermion operators is $\Delta_{2f}=d$.
This aspect recently had also been applied to study far-away-from-equilibrium quantum conformal dynamics~\cite{Maki19}.
In 3D, $\Delta_{4f} \neq 2\Delta_{2f}$
implying strong correlations in pairing channels.
By contrast, for {\em free fermions} with $\tilde{g}=0$, correlation functions shall be given by $\Delta_{4f}=2\Delta_{2f}=2d$ as anticipated; in 3D, this leads to $\Delta_{4f}=6$ instead of $4$.
Moreover, for a generic interacting system with $ 0 > \tilde{g} > \tilde{g}^*(=2-d)$,
$\Delta_{4f}=2d$ also represents the long time asymptotic property 
while the short time asymptotic is still  governed by $\Delta_{4f}=4$. 
On the other hand, when $\tilde{g} < \tilde{g}^*(=2-d) < 0$ or in the strong coupling limit, the solutions of RGEs flow into $\tilde{g}=-\infty$ indicating bound state formation. So the four-fermion correlation function in this limit shall be given by $\Delta_{4f}=d$ and in 3D, $\Delta_{4f}=3$ in Eq.~(\ref{4fC}).
At finite $T$, Eq.~(\ref{4fC}) also sets correlations at $t \ll \hbar/T$
while at longer time, Planckian dynamics set in (see Eq.~(\ref{time})).
Thermodynamics and quantum dynamics of such correlated Planckian liquids are the focuses of discussions below.

\begin{figure}
\includegraphics[width=7cm]{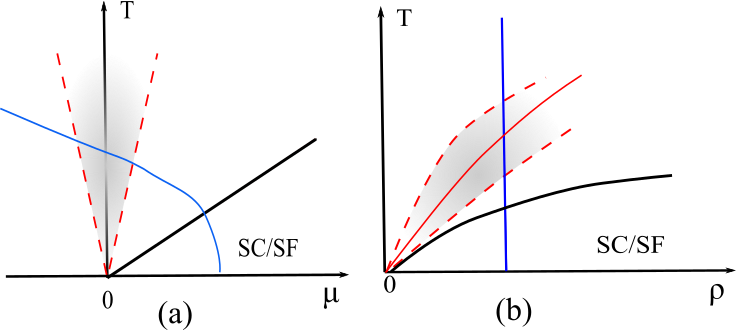}
\centering
\caption{\footnotesize (color online) Regimes (shaded areas bounded by red dashed lines) where Planckian dynamics can be explored (a) in 
the $\mu$-$T$ plane and (b) in the $\rho$-$T$ plane. In a), the region at the lower left corner corresponds to weakly interacting dilute thermal gases and
at lower right is a superfluid or superconducting phase bounded by $T_c (\mu)$, a black solid line.
The blue line in (a) is a typical trajectory followed by fermions interacting with $\tilde{g}^*$ at a given density $\rho$ as temperature $T$ is lowered; it is mapped into a vertical blue line in (b).
Along the red solid line in (b), $\mu(T=T^*, \rho)=0$.}
 \label{Tstar}
\end{figure}

The Planckian limit in this Letter is defined as 3D strongly interacting fermions at finite temperatures $T$ but with $\tilde{\mu}=0$ and $\tilde{g}=\tilde{g}^*=2-d$ as shown in Eqs.~(\ref{RGE}) and (\ref{fp}). Some discussions can also be extended to the vicinity of the fixed point with small $\tilde{\mu}$.

The free energy density $f=F/V$, where the free energy $F=-kT\ln {\mathcal Z}$ can be computed via the partition function ${\mathcal Z}$ and $V$ is the volume.
Using a Hamiltonian ${\mathcal H}(\tilde{g}, \tilde{\mu}; \Lambda)$ defined at a momentum scale $\Lambda$ in Eqs.~(\ref{4fQFT}) and (\ref{RGE}),
one finds that the free energy density $f$ can {\em always} be expressed as
$f(T, \mu, g; \Lambda)=\Lambda^{d+z}  \tilde{f}( \frac{T}{\Lambda^z},\tilde{\mu}(\Lambda),\tilde{g}(\Lambda))$ where $z=2$  defined below Eq.~(\ref{RGE}). 
$\tilde{f}$ shall be a dimensionless but {\em analytical} function of $\tilde{\mu}$, $\tilde{g}$ at any finite temperature $T$~\cite{Z}.
Simply setting $\Lambda=\Lambda_T=T^{1/z}$ and expanding around $\mu/T=0$,
the analysis suggests that near a strong coupling fixed point $\tilde{g}=\tilde{g}^*$ and $\tilde{\mu}=0$, one shall have (see Fig.~\ref{Tstar}), 
\begin{align}
F(T, V, \mu; \tilde{g}=\tilde{g}^*)\approx-V A(\tilde{g}^*) T^{1+\frac{d}{z}} \left(1+ \alpha(\tilde{g}^*) \frac{\mu}{T} \right).
\label{EQ}
\end{align} 
where $A(\tilde{g}^*)$ and $\alpha(\tilde{g}^*)$ are two universal dimensionless quantities depending on the spatial dimension $d$ only.
Exact values can be computed using field theory techniques that  we will return to later when discussing practical detection of Planckian liquids.
The limit of large negative $\mu/T \ll -1$ describes a dilute thermal gas and the limit of $\mu/T \gg 1$ represents a low-temperature superfluid
phase. 

\begin{figure}
\includegraphics[width=8cm]{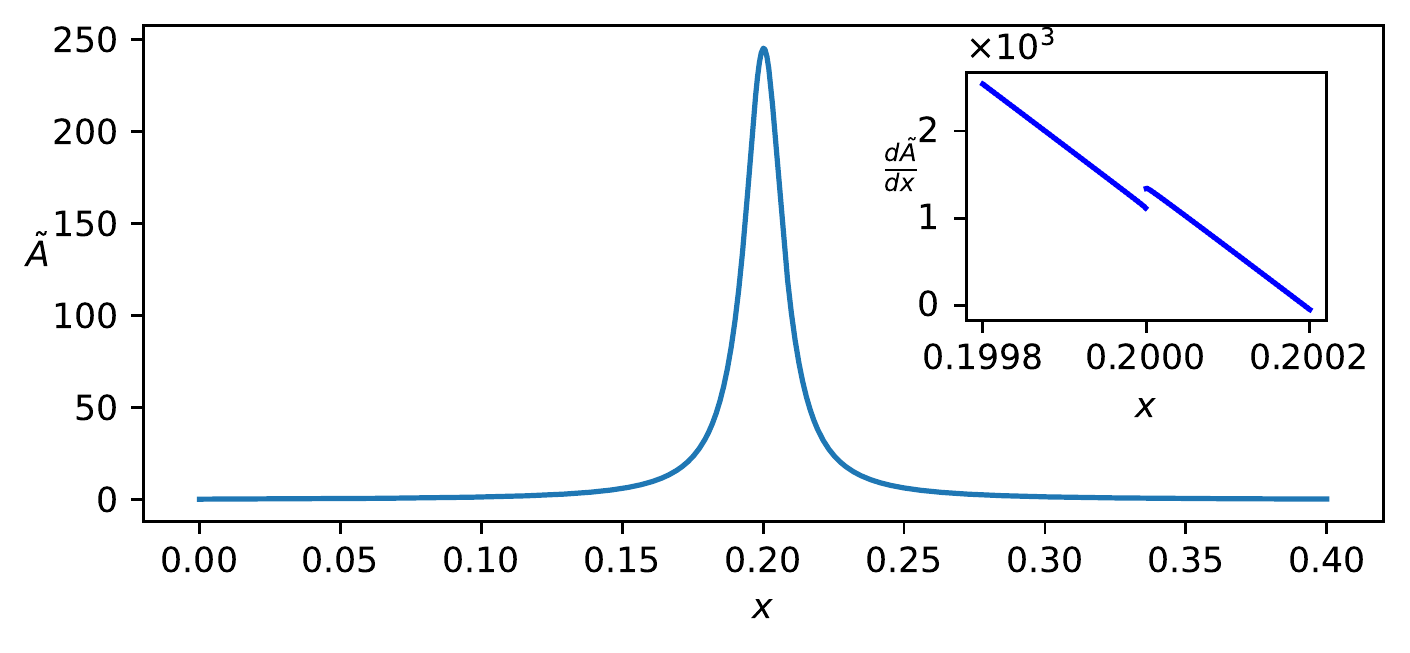}
\centering
\caption{\footnotesize Universal dimensionless spectral function $\tilde{A}(x,y;\epsilon)$ in Eq.~(\ref{SW}) with $\epsilon = 0.1$ and $y=\epsilon_k/T = 0.2$.
Inset shows a kink structure appearing in the leading order of $\epsilon$-expansion.}
 \label{SpecFunc}
\end{figure}

Again we have used $z$, the temporal scaling exponent to represent general results and we set $z=2$ for all discussions in this Letter.  
As $F$ is an analytical function of $\mu$ near $\mu/T=0$,
using general thermodynamic relations one can easily verify that exactly at $\tilde{\mu}=0$, pressure $P(T)$, internal energy density $u(T)$, fermion number density
$\rho(T)$ and entropy density $s(T)$ are uniquely related via the following identities;
\begin{eqnarray}
s(T)&=&
\frac{d+z}{z} A(\tilde{g}^*) T^{\frac{d}{z}};\nonumber  \\
u(T)& =&
\frac{d}{d+z} T s, 
P(T) =
\frac{z}{d+z} T s, \nonumber \\
f(T)  &=&-\frac{z}{d+z} T s,  
\rho(T) = 
\alpha({\tilde{g}^*}) \frac{z}{d+z} s.
\label{TD}
\end{eqnarray}

One notices that a Planckian liquid fundamentally is an {\em entropy liquid} of strongly interacting fermions as all thermodynamic quantities and the equation of states (in units of $T$) are fully described by the entropy density $s(T)$.
Strictly speaking, the identities in Eq.~(\ref{TD}) are only valid at a fixed point and break down whenever $\mu$ is finite or $\tilde{g}\neq \tilde{g}^*$ as in a generic interacting state.
However, practically as far as $\mu \ll T$, the above relations still hold approximately with deviations of the order of $O(\mu/T)$. 
For Planckian liquids we are interested in, we apply a $(2+\epsilon)$-expansion near two spatial dimensions and obtain 
\begin{eqnarray}
A (\tilde{g}^*) &=& \frac{\pi}{12} (1 - 0.565 \epsilon + O(\epsilon^2)), \nonumber \\
\alpha (\tilde{g}^*) &=& \frac{12 \ln 2}{\pi^2}(1+0.261 \epsilon + O(\epsilon^2)).
\label{CO1}
\end{eqnarray}

Dynamics in Planckian liquids we are turning to are unique to strong coupling fixed points, drastically different from dynamics of weakly interacting fermions.
We first focus on the self-energy, $\Sigma^R (\omega, {\bf k})$ and the corresponding spectral weight $A(\omega, {\bf k})=-2 \text{Im} G^R(\omega, {\bf k})$ in the low-energy limit.
Following a general procedure, we identify $\epsilon^*_{\bf k}$, the renormalized particle energy with momentum ${\bf k}$ by setting $\omega -\epsilon^*_{\bf k} -\text{Re} \Sigma^R(\omega=\epsilon^*_{\bf k}, {\bf k})=0$ where the spectral function peaks in the frequency domain.
We next obtain the interaction time, 
$\frac{1}{\tau_{\bf k}} = - Z_{\bf k} \text{Im} \Sigma(\omega=\epsilon^*_{\bf k}, {\bf k})$. Here
$Z_{{\bf k}} =[1-\frac{ \partial \text{Re} \Sigma^R}{ \partial \omega} |_{\omega=\epsilon^*_{\bf k}}]^{-1}$  is a quantity defined near the peak frequency at
$\omega=\epsilon^*_{k}$~\cite{Z_k} and $\frac{1}{{\tau}_{\bf k}}$ specifies the width of the peak.

We have calculated the spectral function $A(\omega, {\bf k})$ in $2+ \epsilon$ dimensions.
Diagrammatically, the leading order contribution to $\text{Im} \Sigma$ comes from a two-loop diagram.
Up to the second order of $\epsilon^2$,
calculations reveal the following structure in the low-frequency limit where $\omega, \epsilon_{\bf k}={\bf k}^2/2 \ll T$~\cite{SE}.

\begin{align}
  \label{SE}
  \text{Re} \Sigma (\omega, \boldsymbol{k}) & = \epsilon^2 T \left( B_1 \frac{\epsilon_{k}}{T} + B_2 \frac{\omega - \epsilon_{k}}{T} \ln \frac{| \omega - \epsilon_{k}|}{T} \right) \nonumber \\
  \text{Im} \Sigma (\omega, \boldsymbol{k}) & = \epsilon^2 T \left( C_0 + C_1 \frac{\epsilon_{k}}{T} + C_2 \frac{\omega - \epsilon_{k}}{T} \right)
\end{align}
with 
\begin{eqnarray}
B_1 &=& 0.091, B_2 = -1/4, C_0 = -0.84,  \nonumber \\
C_1 &=& 0.12, C_2 =0.18 \theta(\omega-\epsilon_k)- 0.61 \theta(\epsilon_k - \omega).
\label{CO2}
\end{eqnarray}
Note Eq.~(\ref{SE}) is not a function of $\omega -\epsilon_{k}$~\cite{chem}.

As a result of Planckian dynamics,
we can define the dimensionless spectral function $\tilde{A}(\frac{\omega}{T}, \frac{\epsilon_k}{T};\epsilon)=A(\omega, {\bf k}) T$. $\tilde{A}(x,y;\epsilon)$ only depends on $d-2=\epsilon$ but is {\em entirely  independent of $T$}.
Following Eq.~(\ref{SE}) when $x, y \ll 1$, we have
\begin{widetext}
\begin{eqnarray}
  \tilde{A}(x,y; \epsilon) =  \frac{2 \epsilon^{2} [C_0 + C_1 y + C_2 (x-y)]}{[x-y-\epsilon^2( B_1 y + B_2 (x-y) \ln |x-y|)]^2 + \epsilon^4 [C_0 + C_1 y + C_2 (x-y)]^2}.
\label{SW}
\end{eqnarray}
\end{widetext}
Eq.~(\ref{SW}) is a universal function of $x$ and $y$ with all coefficients listed in Eq.~(\ref{CO2}). Its dependence on $x$ with a fixed $y$ is shown in Fig.~\ref{SpecFunc}.
And in the low-energy limit,

 \begin{eqnarray}
{\epsilon^*_{\bf k}} &=& \epsilon_{\bf k} (1+B_1 \epsilon^2)+...,\nonumber \\
\frac{\hbar}{\tau_{\bf k}} & =&
 \epsilon^2 T \left(0.84 -0.12 \frac{{\epsilon_{\bf k}}}{T} \right) +... 
\label{time}
 \end{eqnarray}
By extrapolating to $\epsilon=1$, one gets an estimate for 3D systems. In $d=3$, Eq.~(\ref{time}) indicates a Planckian time scale, $0.84 \hbar/T$ near ${\bf k}=0$.
By contrast, in the weakly interacting limit, $\frac{1}{\tau_{\bf k}} \sim g^2 T^2$, and is quadratic in $T$. 
However, for higher energy particles with $\epsilon_{\bf k} \gg T$, we find that $\frac{1}{\tau_{\bf k}} \sim \epsilon^2 T (\frac{T}{\epsilon_{\bf k}})^{\epsilon/2}$ implying a distinctly different ultraviolet limit.
Eqs.~(\ref{TD})--(\ref{time}) and Eq.~(\ref{mu}) below are the main results of this article.

At last, as a Planckian liquid is fully determined by entropy density $s(T)$ and Planckian time $\hbar/T$, naturally its quantum dynamics is uniquely set with $\hbar$.
For example, shear viscosity at a strong coupling fixed point where $\tilde{\mu}=0$ shall have the form of $\zeta(T, \tilde{g}^*) = D(\tilde{g}^*) T^{\frac{d}{z}}$,
where again $z=2$ and $D$ is a function of $\tilde{g}^*=2-d$, or spatial dimension $d$.
Note that $\zeta(T)/s(T)$ is independent of $T$, also a {\em hallmark signature} of Ads/CFT quantum dynamics with a blackhole~\cite{Kovtun05}.
At $2+\epsilon$ dimensions, we obtain $D \approx 0.156 \epsilon^{-2}$~\cite{eta}.

So far we have outlined general properties of the high-dimensional Planckian liquids.
Most of physics discussed can be simulated by connecting many-body states with a particle reservoir where the chemical potential is fixed exactly
at $\mu=0$. If it is practically challenging to achieve such a realization, we suggest an alternative route for a state with a fixed density.

Our proposal relies on a simple thermodynamic relation between the Planckian liquid so far discussed and a fixed-density
quantum many-body system. The chemical potential in a fixed density $\rho$ and temperature $T$ is a function of both parameters. So if we only sample data from  
low density fermions at a given temperature $T^*$ where
\begin{eqnarray}
\mu(\rho, T=T^*; g^*)=0, \text{ or    }  T^*=\left( \frac{\rho}{\alpha(\tilde{g}^*) A(\tilde{g}^*)} \right)^{2/3},
\label{mu}
\end{eqnarray}
we effectively single out a Planckian liquid with thermodynamics and dynamics given in Eq.~(\ref{TD}),(\ref{SW}) defined at temperature $T^*$.
The second equation in Eq.~(\ref{mu}) follows 
the relation between $\rho$ and $T$ in Eq.~(\ref{TD}) that is only valid for Planckian liquids at $\mu=0$.
The topology of Fig.~\ref{Tstar}(a) illustrates that a Planckian liquid can not be superconducting or superfluid as
$T^*$ is always above superconducting or superfluid transition temperature $T_c$.
Extrapolating Eq.~(\ref{mu}) to $\epsilon = 1$, one gets $T^{*} \approx 0.73 T_F$, while latest experiment shows that $T_{c} \approx 0.17 T_F$ at unitarity.
Fig.~\ref{Tstar} explicitly indicates  a unique role of $T^*$ in quantum dynamics.

\begin{acknowledgments}
This project is in part supported by NSERC, Canada under a Discovery grant under RGPIN-2020-07070 and National Natural Science Foundation of China (Grant No. 11804376).
FZ wants to thank L. Taillefer, J. Zaanen for discussions on Planckian transport in cuprates, and M. Ku, J. Maki for discussions on hydrodynamics.
\end{acknowledgments}


\begin{thebibliography}{References}
\bibitem{mu} Strictly speaking, Eq.~(\ref{RGE}) is valid for $\Lambda^z > \mu$. 
When $\Lambda^z \ll \mu$, $\beta_g = \tilde{g}^2$ implying a Cooper instability.
Here we have set $\mu=0$ first when deriving it.
\bibitem{Z} $Z$ does not play a role in this discussion as $Z=1$ at $T=0$ so for simplicity we have muted it in Eq.~(\ref{EQ}).
\bibitem{chem} The chemical potential $\mu (T) =\mu (T=0) -\epsilon B_0 T$ and has been fixed to be zero. Here
$B_0 = - \ln 2 + \epsilon 0.34$.
\bibitem{eta}
We contrast the shear viscosity obtained for a Planckian liquid with known results of a weakly interacting counterpart (but with $\mu=0$); 
in the latter case, one can verify that in 3D,
$\zeta \sim \frac{1}{g^2}  T^{1/2}$ and $\zeta(T)/s(T) \sim \frac{1}{g^2 T} \gg 1$ which is divergent at low $T$ limit. As a result of weaker interactions, when $T$ approaches zero,
both $\zeta(T)$ and $\zeta(T)/s(T)$ are much higher than their counterparts in a 3D Planckian liquid.
\end{thebibliography}
\end{document}